# The Mechanical Paul Trap:
# Introducing the Concept of Ion Trapping

Sebastian Kilde Lofgren[*], Ricardo Méndez Fragoso[**], Jonathan Weidow[***], Jonas Enger[*]

Nobel laureate Wolfgang Paul showed, back in the 1950s, that charged particles can be trapped using alternating electric fields[1]. This technique is commonly referred to as Paul traps or radiofrequency traps (RF-traps) and is used in various areas of modern physics. This paper presents a 3D-printed mechanical Paul trap, a naïve simulation of the system in Python, and student investigations. The files for the 3D-printed trap are available for download and print[i], and the code for the simulation is available to run and tinker with[ii].

The mechanical Paul trap (*Fig. 1A*) can be used as a demonstration tool or an experimental setup to simulate how a real Paul trap works or to investigate interesting physical phenomena[2–5]. Previous work has mainly focused on higher education, and here we present investigations that can be incorporated into lower parts of the educational system. Proposed investigations can either be done in isolation or together to have the student(s) experience the physical phenomena in the mechanical Paul trap and visualize how a real Paul trap works.

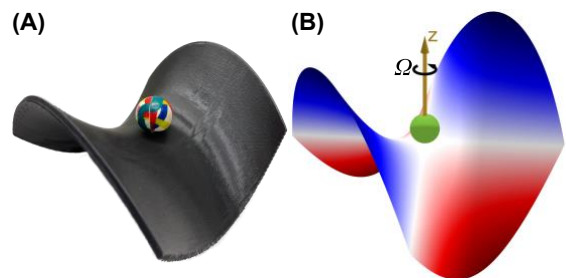

*Fig. 1:* **(A)** *3D-printed mechanical Paul trap.* **(B)** *A particle in the quadrupole field. The potential of the field is illustrated as both the height of the saddle and the color scale that goes from dark red (large attractive force) to dark blue (large repulsive force).*

## The Mechanical Paul Trap

A linear Paul trap confines particles by having four metal rods connected to an AC source, where opposite rods have the same polarity and continuously change their polarity with some frequency $\Omega$, as seen in *Fig. 2*.

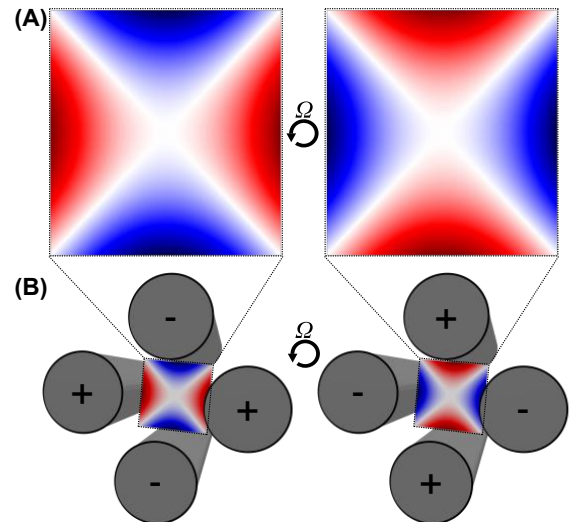

*Fig. 2:* **(A)** *Two configurations of the potential field in a linear Paul trap. The normalized potential goes from -1 (dark blue) to 1 (dark red). The field flips from left to right configuration during its rotation driven by the angular frequency $\Omega$. The potential is zero along the diagonals, creating a saddle point in the middle.* **(B)** *A linear Paul trap using 4 charged rods.*

Moving from a two-dimensional representation to a three-dimensional one makes visualizing the field strength less abstract. *Fig. 1B* shows a saddle-looking shape with a particle in the middle. In this picture, the field strength is represented as the height of the z-axis. However, the field is not visible and thus becomes a rather abstract concept to communicate to students. This raises critical pedagogical questions about visualizing and making the idea more tangible for students. Also, why do we even need a varying electric field to trap a particle? One way is to take a pure mathematical route and let that be the representational tool to understand the physical situation. Another method can be finding some analog that students can see and relate to. This is where the mechanical Paul trap takes the stage, as it did during Wolfgang Paul's Nobel lecture[1]. Since then, others have constructed and studied the mechanical Paul trap for demonstration purposes or laboratory exercises [4–6].

[*] University of Gothenburg, Sweden
[**] Universidad Nacional Autónoma de México, Mexico
[***] Chalmers University of Technology, Sweden
Corresponding Author: Sebastian Kilde Lofgren, sebastian.lofgren@physics.gu.se

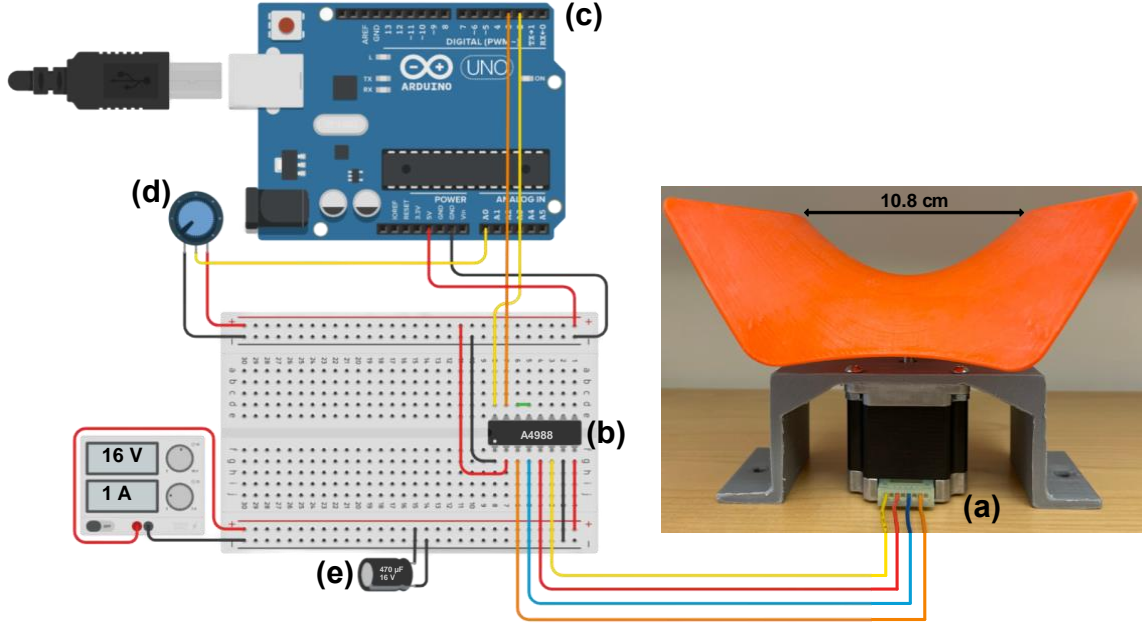

*Fig. 3: Schematics for controlling the mechanical Paul trap with an Arduino. Using a NEMA 23 stepper motor (a) to drive the saddle surface requires a motor driver (b). The Arduino (c) is connected to a computer for displaying the frequency. The frequency is controlled by a potentiometer (d). A capacitor (e) protects the motor from current spikes.*

However, the mechanical Paul trap is not a trivial analog because it has some crucial differences from a real Paul trap. Two such significant differences are the effect of rolling and friction, which can lead to trapping balls on various rotating surfaces[7–10]. The effects in a mechanical Paul trap have been discussed as simulating adding a magnetic field to the Paul trap, giving rise to a Lorentz force[3,6,11]. There have been several proposals for alternative models that can be used for demonstration purposes, such as a spring-loaded inverted pendulum[2], the rotating hoop trap[12], or a large-scale electrical quadrupole trap[13]. The first two examples solve some apparent problems with the mechanical Paul trap model but also remove some pedagogical strengths, such as the potential shape's physical appearance. The last example lets students see particles trapped in a varying electric field. However, since it uses high voltage, the equipment is dangerous, and once again, the visual component of the particle-field interaction is obscured. Despite its shortcomings, the focus on the visual aspect of particle-field interaction is why the mechanical Paul trap is further investigated as a pedagogical tool in this paper.

A mechanical Paul trap can be constructed, as described by Rueckner et al.[5] or 3D-printed. There are several advantages of using an automated process like 3D printing, such as low cost, high precision, and easy modification. The proposed trap design only requires a few components that cannot be 3D-printed: a motor, an Arduino and some basic components, and a power source. See **Fig. 3** for schematics[iii].

### Theory

For a real Paul trap, we can express the varying electric potential as

$$V(t) = \frac{V_0}{2R^2}(x^2 - y^2)\cos(\Omega t) \quad (1)$$

where $V_0$ is the voltage of an AC source, $R$ is half the distance between two opposing rods (**Fig. 1**), $\Omega$ is the angular frequency of the AC source, $t$ the time, and $x$ and $y$ are cartesian coordinates.

For the mechanical Paul trap, in the frictionless case, we can construct the gravitational potential as

$$U_g(t) = \frac{mgh}{R^2}[(x^2 - y^2)\cos(2\Omega t) + 2xy\sin(2\Omega t)] \quad (2)$$

where $m$ is the mass of the trapped particle, $g$ the gravitational acceleration, $\Omega$ the angular rotational frequency of the saddle surface, $R$ the radius of the saddle, $h$ the maximum height of the surface at a distance $r$ from the center, $t$ the time, and $x$ and $y$ are cartesian coordinates.

Equations (1) and (2) look alike, apart from a sinusoidal cross-dependent term for the mechanical trap. This term shows that the potential in the mechanical Paul trap is bounded[4], which must be true for a solid surface. The gravitational potential and the initial conditions dictate the motion of a "particle" in the trap. In equation (2), the shape of the trap, given by $\frac{h}{R^2}$, and the angular frequency $\Omega$ are controllable terms that affect the "particle's" path. In the case of the real Paul trap, the shape of the trap and the frequency at which the electric potential is changed are terms affecting particle paths as well. Thus, to use the mechanical Paul trap to teach how real Paul traps work, the shape and frequency can be relevant factors to investigate.

### Pedagogical framework

The theoretical framework that served as a guiding principle when developing the exercises is variation theory proposed by Marton[14,15]. A fundamental question to ask is what students need to have learned to succeed in a particular task[14]? I.e., what are the critical aspects when learning something? According to variation theory, learning something can be thought of as *learning to see*. Learning is about being able to discern features of the world, and the proposed process of going from *not seeing* to *seeing* can be separated into three stages that Marton[14] refers to as

1. **Contrast**, where different aspects of a concept of phenomena take turns being varied or invariant. This can be done via induction, where the studied aspect is invariant, or via contrast, where the studied aspect is varied.
2. **Generalization** lets the discerned feature in the previous stage be invariant and allows the student to vary other aspects.
3. **Fusion** requires that the student experiences variations of multiple focused aspects and puts different aspects together to understand the object of learning more deeply.

### Student investigation 1: Physical properties of a saddle point

In Swedish upper secondary mathematics, students experience the concept of saddle points in the abstract world of sketching function graphs using derivatives. A saddle point exhibits the phenomena of having a derivative of zero, but when moving just a small distance away from that point, the derivatives point toward the saddle point in some directions and away in others. For example, the function $z = \frac{1}{a}(x^2 - y^2)$ constructs the surface of the mechanical Paul trap (*Fig. 1*). Using the mechanical Paul trap, one can intuitively understand the concept of saddle points by trying to place a ball on the trap carefully and have it not fall off when the trap is stationary. It may require some trial and error, but it is doable. To guide students, the teacher can ask follow-up questions: What does it take for the ball not to roll off? Why does the ball tend to roll off the surface?

This simple experiment highlights one crucial property of saddle points. Namely, it is possible to have something placed on such a point, but it is not stable.

### Student investigation 2: Varying frequency

The mechanical Paul trap has a frequency region where, if a ball is placed close to the middle with some care, it will be trapped for some time. The purpose of varying the rotation frequency of the trap is to guide students to discern rotation frequency as a critical aspect through contrast.

We propose the following procedure to discover the importance of rotation frequency in the mechanical Paul trap:

a) Place a ball, for example, a glass bead with a diameter of 22 mm, on the saddle surface using the following rotational frequencies: 2 Hz and 3.5 Hz. Repeat the experiment at least ten times for each frequency and note the trapping time for each trial. Is it possible to catch the ball at any of the frequencies?

The trapping time can be defined as the time from letting go of the ball until the ball leaves the trap. The ball can be counted as trapped if the trapping time exceeds, for example, 10 seconds.

b) Using a computer simulation, investigate if a "particle" can be trapped using the following rotational frequencies: 2 Hz and 3.5 Hz, and the starting position $x = 0$ mm and $y = 8$ mm.

The frequencies are chosen so students will experience one region where a ball cannot be trapped and one where trapping is possible. For example, when trying to trap the glass bead at 3.5 Hz, it is possible to trap it for at least 10 seconds. However, due to the chaotic nature of the mechanical Paul trap, the trapping time can fluctuate between 1 – 13 seconds (*Fig. 4*).

### Student investigation 3: Threshold frequencies

There is a range of possible frequencies to trap a ball, and finding the lower bound could help highlight this feature. The lower bound is the threshold frequency, which we define as where the trapping time increases significantly.

For this investigation, let the students choose their approach and have them vary the size of the ball (for example, between 10 – 45 mm in diameter). Some groups may rely heavily on computer simulation, and others do not use it. Moreover, when using the mechanical Paul trap, one can find the threshold frequency in a couple of different ways: (1) starting with a frequency where it is easy to trap and with the ball stable in the trap, successively lowering the frequency until dropping the ball; (2) starting with a low frequency and systematically increasing the frequency in small increments. The threshold frequency should become somewhat apparent if plotting the trapping time at various frequencies, as in *Fig. 4*.

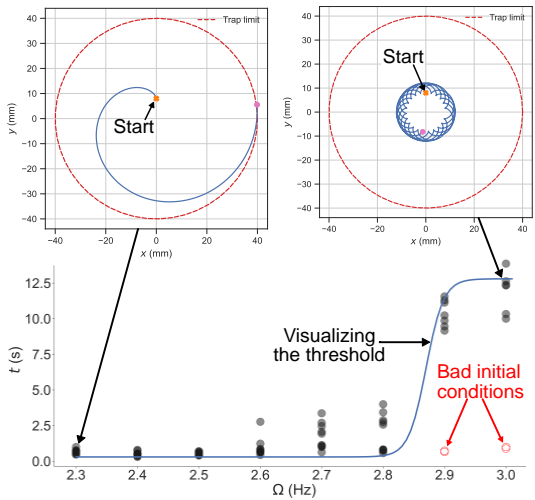

Fig. 4: Finding the threshold frequency for a glass bead with a diameter of 22 mm. The trapping time increases significantly from 2.8 to 2.9 Hz, usually by at least one order of magnitude. The inclusion of bad initial conditions here highlights the sensitivity of initial conditions. With the simulation, this shift is observable as the particle being either trapped indefinitely or not by observing a few rotations.

The act of changing balls affects the system in various ways. If the mass or surface of the ball is changed, then friction is changed. However, small changes in friction only appear to affect trapping time and not the threshold frequency. Generally, it is suggested to have low friction in the system to increase the trapping time. When the ball's size changes, the threshold frequency changes since the effective potential changes. The effective potential can be considered the allowed path of a ball's center of mass in the trap. The simulation can help highlight this by having the ball as a frictionless point particle.

## Conclusions

Our proposed investigations allow students to explore the concept of trapping particles using varying fields. Furthermore, the abstract notion of field strength can be explained using multiple representations by having a mechanical analog to the Paul trap and a simulation. This opens more possibilities for students to grasp the concept through variation.

## Acknowledgment

The authors would like to thank Prof. Dag Hanstorp for introducing the mechanical Paul trap to the group and the anonymous reviewers for their help in improving the paper.

R.M.-F. acknowledges the financial support from The Wenner-Gren Foundations and the grants DGAPA-PASPA, PAPIIT IN-112721 and PAPIME PE-103021.

---

[i] Link to stl-files ready to be 3D-printed:
https://www.thingiverse.com/thing:5249350

[ii] Link to a Google Colaboratory notebook with the simulation, ready to run directly from the browser:
https://colab.research.google.com/drive/1j8PF0aVwbNHfwCPKdJB4rMXY4U9XapK1?usp=sharing

[iii] Link to the Arduino code:
https://github.com/SebKilL91/Mechanical-Paul-trap